\documentclass[cits]{PoS}\usepackage{amsmath,amsbsy,cite}

\title{Flavour-Exotic Tetraquarks in Large-$N_{\rm c}$ QCD:\\Do
They Exist?}\ShortTitle{Flavour-Exotic Tetraquarks in
Large-$N_{\rm c}$ QCD: Do They Exist?} 
\author{Wolfgang Lucha\\Institute for High Energy Physics, Austrian
Academy of Sciences, Nikolsdorfergasse 18,\\A-1050 Vienna,
Austria\\E-mail: \email{Wolfgang.Lucha@oeaw.ac.at}}
\author{\speaker{Dmitri Melikhov}\\Institute for High Energy
Physics, Austrian Academy of Sciences, Nikolsdorfergasse
18,\\A-1050 Vienna, Austria, and\\D.~V.~Skobeltsyn Institute of
Nuclear Physics, M.~V.~Lomonosov Moscow State University,\\119991,
Moscow, Russia, and\\Faculty of Physics, University of Vienna,
Boltzmanngasse 5, A-1090 Vienna, Austria\\E-mail:
\email{dmitri\_melikhov@gmx.de}}\author{Hagop Sazdjian\\Institut
de Physique Nucl\'eaire, CNRS-IN2P3, Universit\'e Paris-Sud,
Universit\'e Paris-Saclay,\\91405 Orsay Cedex, France\\E-mail:
\email{sazdjian@ipno.in2p3.fr}}\abstract{Flavour-exotic tetraquark
mesons, by definition bound states of two quarks and two
antiquarks of four mutually different quark flavours, are, for
given Lorentz features, subject to two incompatible constraints:
On the one hand, within quantum chromodynamics a formation of
compact tetraquark states is most easily envisaged by merging two
colour-antisymmetric two-quark clusters, a diquark and an
antidiquark. This path, however, leads to merely a single
tetraquark state of chosen Lorentz characteristics. On the other
hand, in the limit of the number of colour degrees of freedom
growing beyond bounds, internal consistency at leading order calls
for the presence of (at least) two of such tetraquark states of
\emph{identical\/} quark-flavour composition. The failure of
attempts to reconcile these two contradictory insights suggests
the nonexistence of compact flavour-exotic tetraquark~mesons.}

\FullConference{European Physical Society Conference on High
Energy Physics -- EPS-HEP2019\\10--17 July, 2019\\Ghent, Belgium}

\begin{document}\section{Simplified Look at Tetraquarks by Quantum
Chromodynamics in Large-$N_{\rm c}$ Limit}Quantum chromodynamics
(QCD), the quantum field theory governing the strong interactions,
supports as admissible bound states of its quark and gluon degrees
of freedom not only conventional hadrons (i.e., the
quark--antiquark mesons and three-quark baryons) but also
multiquark states (such as tetraquarks, pentaquarks, hexaquarks)
and hadrons hybridized by gluon excitations. The first and
foremost boundary condition to each variant of exotic construction
is: it has to form a colour singlet. We are interested in
tetraquark mesons exhibiting flavour quantum numbers
$a,b,c,d\in\{u,d,s,c,b\}$,\begin{equation}T=[\overline
q_a\,q_b\,\overline q_c\,q_d]\ ,\label{fe}\end{equation}compact
states tightly binding two quarks $q_b,q_d$ and two antiquarks
$\overline q_a,\overline q_c$, of masses $m_a,m_b,m_c,m_d$.

Our goal is to derive information on the main features of
tetraquark mesons \cite{TQN1,TQP1,TQN2,TQN3,TQP2,ESR} by
inspection of their contributions, in form of poles, to the
amplitudes of the scattering of two ordinary mesons of momenta
$p_1$ and $p_2$ into two ordinary mesons of momenta $p_1'$ and
$p_2'$. Supposing these four ordinary mesons to be created from
the vacuum by adequate interpolating operators or currents, we do
this by investigating Green functions of four quark-bilinear
currents, here generically symbolized by $j$. For the meson
$M_{\overline ab}$ composed of an antiquark $\overline q_a$ and a
quark $q_b$ with associated field operators~$\overline q_a(x)$ and
$q_b(x)$, its coupling strength $f_{M_{\overline ab}}$ to any
appropriate quark-bilinear operator $j_{\overline ab}(x)$, upon
notationally suppressing immaterial reference to Lorentz nature of
generic form $j_{\overline ab}(x)\equiv\overline q_a(x)\,q_b(x)$,
is given~by$$f_{M_{\overline ab}}\equiv\langle0|j_{\overline
ab}(0)|M_{\overline ab}\rangle\ .$$

The primary task in that undertaking is to formulate a criterion
that allows for the unambiguous identification of those
contributions to the above scattering amplitudes that
encapsulate~information about tetraquark mesons \cite{TQN1}.
Feynman diagrams passing this criterion are
labelled~\mbox{\emph{tetraquark-phile\/}:}\begin{quote}A Feynman
diagram is tetraquark-phile \cite{TQP1,TQP2} if, as function of
the Mandelstam variable$$s\equiv(p_1+p_2)^2=(p_1'+p_2')^2\ ,$$it
depends \emph{non-polynomially\/} on $s$ and exhibits a branch cut
starting at the \emph{branch~point\/}~$\hat s$ defined by the
square of the sum of the masses of the involved bound-state
constituents,$$\hat s=(m_a+m_b+m_c+m_d)^2\ ,$$possibly
contributing to a \emph{tetraquark pole\/} by support of a
four-quark intermediate state.\end{quote}The (non-) existence of
such a branch cut can be decided by application of the Landau
equations \cite{LDL}.

In the present context, considerable profit will be drawn from
allowing the number $N_{\rm c}$ of~colour degrees of freedom of
QCD to become arbitrarily large and harvesting any finding of the
formulated theory, large-$N_{\rm c}$ QCD \cite{GH,EW}, in both its
limit $N_{\rm c}\to\infty$ and its expansion in powers of
$1/N_{\rm c}$ thereabout. Consistency calls for constancy of the
product of $N_{\rm c}$ and the square of the strong coupling
$g_{\rm s}$ \cite{GH},~i.e.,$$\alpha_{\rm s}\equiv\frac{g_{\rm
s}^2}{4\pi}=O(N_{\rm c}^{-1})\xrightarrow[N_{\rm c}\to\infty]{}0\
.$$An immediate implication of this is the large-$N_{\rm c}$
behaviour of all ordinary-meson couplings $f_{M_{\overline ab}}$
\cite{EW},$$f_{M_{\overline ab}}=O(\sqrt{N_{\rm c}})\qquad
\mbox{for}\quad N_{\rm c}\to\infty\ .$$Compared with its point of
origin, large-$N_{\rm c}$ QCD is of a significantly lesser
complexity; this enabled (at least, qualitative) analyses of
different features of hadrons, notably, of multiquark states
\cite{SW,KP,CL,MPR1,MPR2}.

\section{\emph{Flavour-Exotic\/} Tetraquark Poles in Correlators:
Large-$N_{\rm c}$ Leading Contributions}Evidently, our next move
must be to define the tetraquarks' flavour content we are
interested in. Table~1 of Ref.~\cite{TQN3} compiles the
conceivable quark-flavour combinations. From this list,
particularly tantalizing to us seems to be the case of
flavour-exotic tetraquarks, where the flavours $a,b,c,d$ of~the
two quarks and two antiquarks forming any such meson differ from
each other. This case belongs to those configurations that betray,
beyond all doubt, the \emph{nonconventional\/} nature of some
hadron~state.

We study mere two-ordinary-meson scattering but (for the sake of
completeness) with either~an identical or a differing distribution
of the four available quark flavours to the two ordinary mesons in
initial and final state. As a consequence, we have to take into
account two categories\footnote{In order to prevent confusion,
recall that in Ref.~\cite{TQN1} these sets got named ``direct''
and ``recombination'', respectively.} of correlators:\begin{align}
\mbox{flavour-preserving Green functions}&=\left\{\langle
T(j_{\bar ab}\,j_{\bar cd}\,j^\dag_{\bar ab}\,j^\dag_{\bar
cd})\rangle,\ \langle T(j_{\bar ad}\,j_{\bar cb}\,j^\dag_{\bar
ad}\,j^\dag_{\bar cb})\rangle\right\},\label{fpG}\\
\mbox{flavour-reordering Green functions}&=\left\{\langle
T(j_{\bar ad}\,j_{\bar cb}\,j^\dag_{\bar ab}\,j^\dag_{\bar
cd})\rangle\right\}.\label{frG}\end{align}Any \emph{potential\/}
intermediate-state tetraquark satisfies: if it contributes to
\emph{both\/} of the flavour-retaining correlators (\ref{fpG}),
then it will contribute to the flavour-reshuffling correlator
(\ref{frG}) too, and \emph{vice versa\/}.

\begin{figure}[hbt]\centering
\includegraphics[scale=.49095,clip]{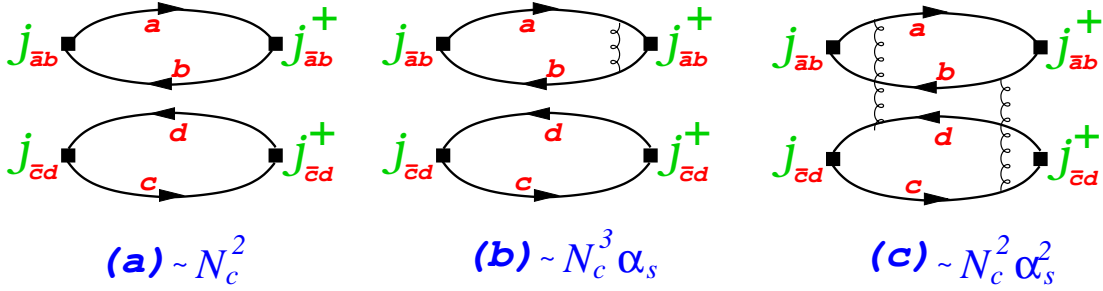}
\caption{Flavour-preserving Green functions $\langle T(j_{\bar
ab}\,j_{\bar cd}\,j^\dag_{\bar ab}\,j^\dag_{\bar cd})\rangle$ of
four quark-bilinear currents~$j$: examples of contributions by, at
order $O(N_{\rm c}^2)=O(N_{\rm c}^3\,\alpha_{\rm s})$ necessarily,
\emph{non-tetraquark-phile\/} Feynman diagrams (a,b) as well as
\emph{tetraquark-phile\/} Feynman diagrams (c), of the
\emph{$N_{\rm c}$-leading tetraquark-phile\/} order $O(N_{\rm
c}^0)=O(N_{\rm c}^2\,\alpha_{\rm s}^2)$.}\label{tpp}\end{figure}

\begin{figure}[hbt]\centering
\includegraphics[scale=.47257,clip]{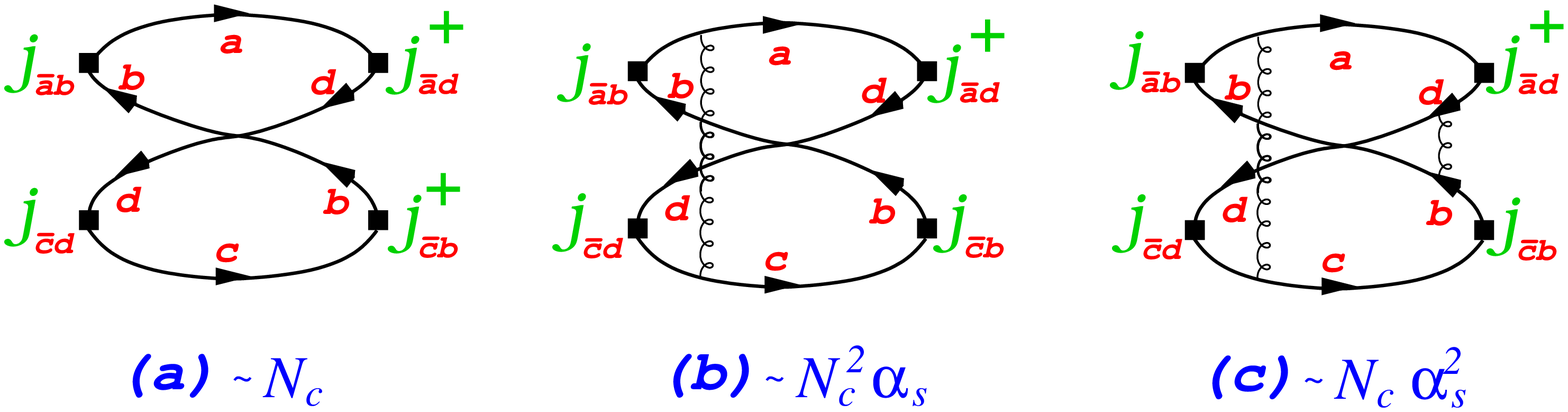}
\caption{Flavour-reordering Green functions $\langle T(j_{\bar
ad}\,j_{\bar cb}\, j^\dag_{\bar ab}\,j^\dag_{\bar cd})\rangle$ of
four quark-bilinear currents~$j$: examples of contributions by [at
order $O(N_{\rm c})=O(N_{\rm c}^2\,\alpha_{\rm s})$ exclusively]
\emph{non-tetraquark-phile\/} Feynman diagrams (a,b) as well as
\emph{tetraquark-phile\/} Feynman diagrams (c) of the $N_{\rm
c}$-leading tetraquark-phile order $O(N_{\rm c}^{-1})=O(N_{\rm
c}\,\alpha_{\rm s}^2)$.}\label{tpr}\end{figure}

\begin{figure}[hbt]\centering
\includegraphics[scale=.49854,clip]{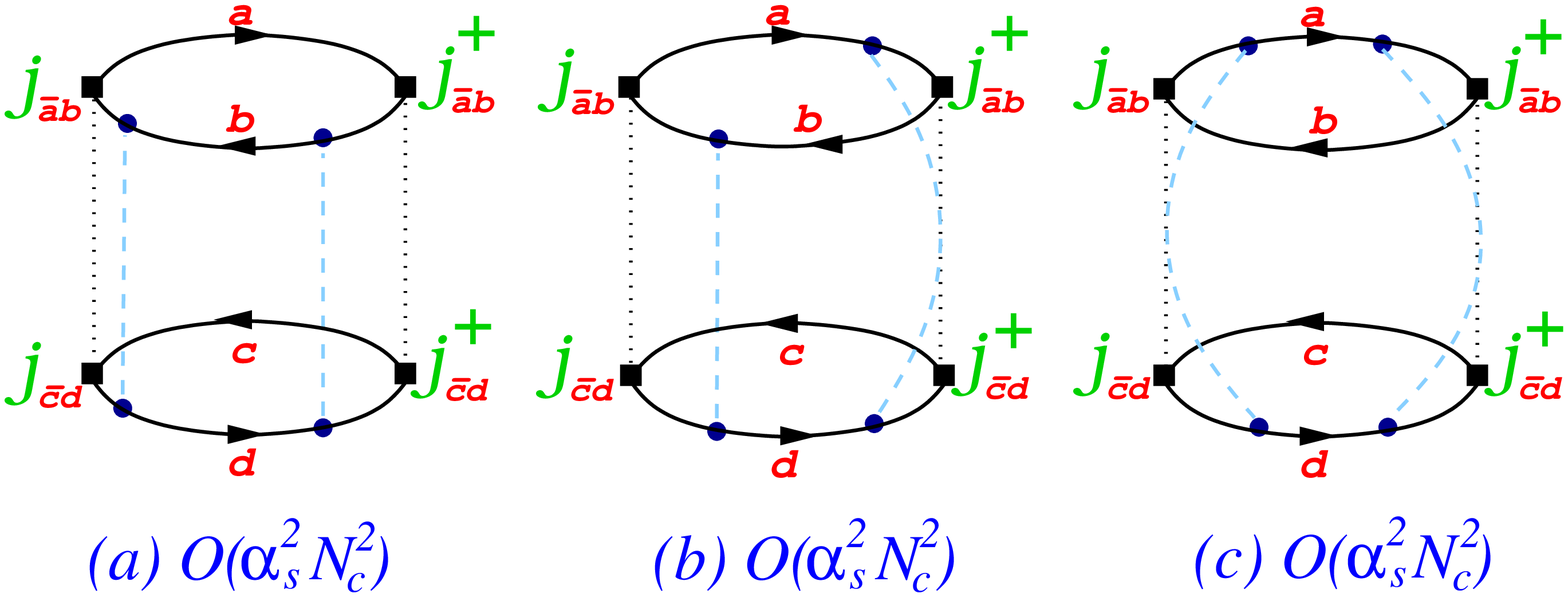}
\includegraphics[scale=.49854,clip]{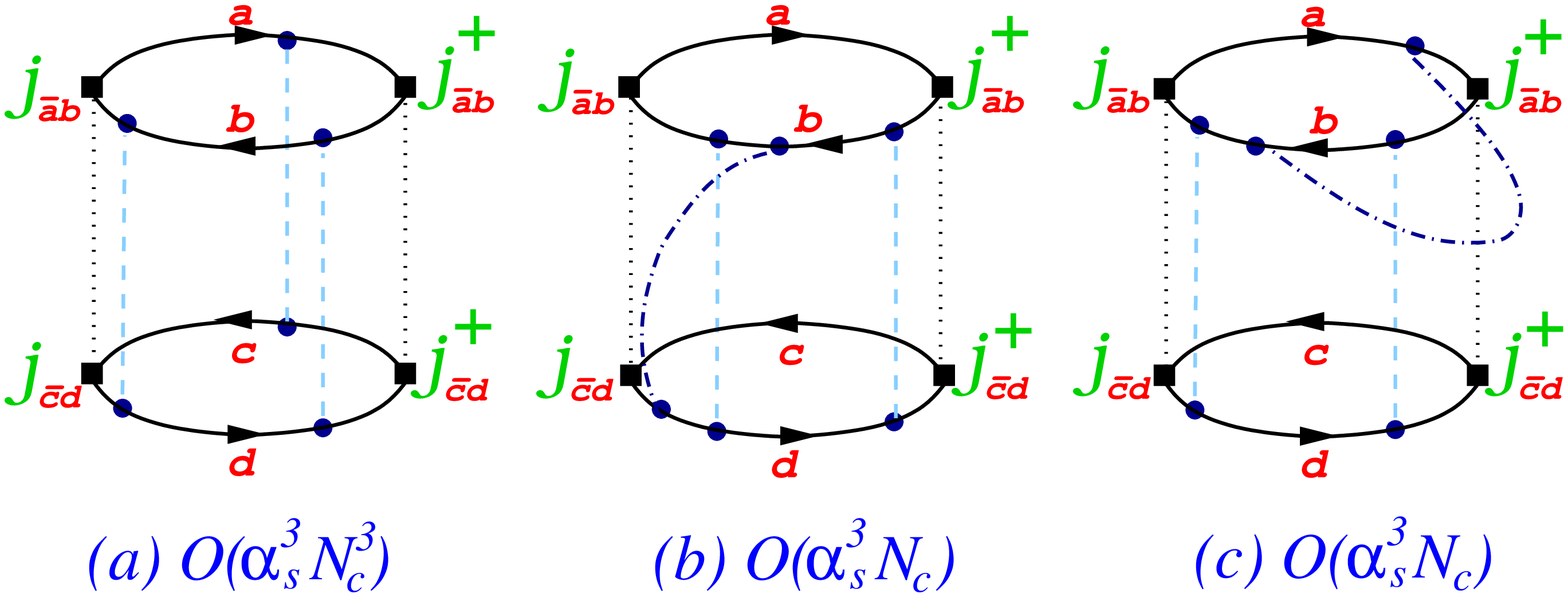}\caption{Cylinder
interpretation (with cylinder surfaces indicated by dotted black
lines) of typical examples of the flavour-retaining
tetraquark-phile Feynman diagrams \cite{TQP2} of (top) $N_{\rm
c}$-leading order $O(\alpha_{\rm s}^2\,N_{\rm c}^2)=O(N_{\rm
c}^0)$, involving two planar gluons (pale blue dashed lines), and
an amendment of Feynman diagram (a) by (bottom) one additional (a)
planar gluon (pale blue dashed lines) or (b,c) nonplanar gluon
(dark-blue dot-dashed lines).}\label{fp(+)}\end{figure}

\begin{figure}[htb]\centering
\includegraphics[scale=.49854,clip]{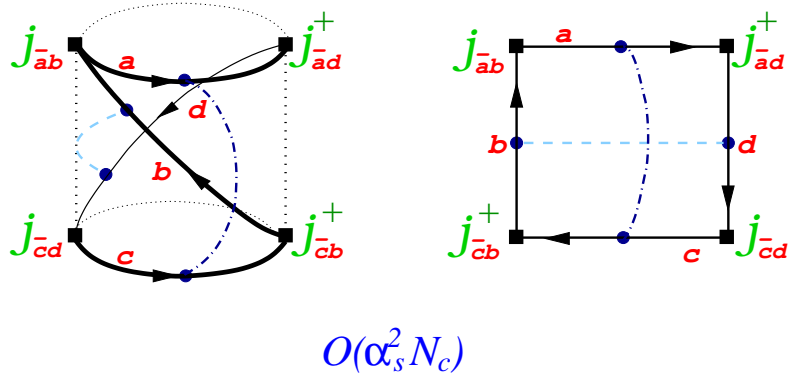}\caption{Generic
example of flavour-reshuffling tetraquark-phile Feynman diagrams
\cite{TQP2} of $N_{\rm c}$-leading~order $O(\alpha_{\rm
s}^2\,N_{\rm c})=O(N_{\rm c}^{-1})$, involving one planar gluon
(pale blue dashed lines) and one nonplanar gluon (dark-blue
dot-dashed lines), in cylinder (left, with the cylinder contours
indicated by dotted black lines) and~in~unfolded (right) ---
therefore the inevitable nonplanarity of one of the exchanged
gluons emphasizing --- interpretation.}\label{fr}\end{figure}

For the elements of the two disjoint sets (\ref{fpG}) and
(\ref{frG}) of Green functions, the Laurent series of the
$1/N_{\rm c}$ expansions of their contributions start at order
$O(N_{\rm c}^2)$ for the flavour-preserving (Fig.~\ref{tpp}) set
(\ref{fpG}) but at order $O(N_{\rm c})$ for the flavour-reordering
(Fig.~\ref{tpr}) set (\ref{frG}), respectively. The corresponding
decisive subsets of \emph{tetraquark-phile\/} contributions prove
to be \cite{TQN1,TQP1,TQN2,TQN3} of maximum order $O(N_{\rm c}^0)$
in the flavour-retaining case (Fig.~\ref{fp(+)}) but of maximum
order $O(N_{\rm c}^{-1})$ in the flavour-reordering case
(Fig.~\ref{fr}). Imagining a tetraquark-phile Feynman diagram in
the form of a cylinder (Fig.~\ref{fp(+)}, bottom row;
Fig.~\ref{fr}, left) facilitates its inspection, in particular,
the crucial distinction of its planar and~nonplanar gluons.
Disregarding (hardly justifiable) adjustments by hand of $N_{\rm
c}$ dependencies leads us to the finding \cite{TQN1}:\begin{quote}
The \emph{$N_{\rm c}$-leading\/} members of the two classes of
\emph{tetraquark-phile\/} Feynman diagrams,~the flavour-preserving
and the flavour-rearranging ones, exhibit a different $N_{\rm c}$
dependence.\end{quote}

A consequence of this outcome is that the implied constraints on
the large-$N_{\rm c}$ limit of the~Green functions cannot be
solved under the assumption of the presence of only a single
tetraquark. Rather, for a selected quark flavour combination one
needs \emph{at least\/} two tetraquarks, discriminable, however,
by the large-$N_{\rm c}$ dependence of their transitions to the
two conceivable pairs of conventional mesons.

For any flavour-exotic quark content, let us focus to the minimal
option, i.e., two tetraquarks~$T_A$ and $T_B$ of masses $m_{T_A}$
and $m_{T_B}$, respectively. Their impact on the correlators
(\ref{fpG}) and (\ref{frG}) yields, in terms of all
tetraquark--two-ordinary-meson transition amplitudes
$A(T_\square\longleftrightarrow M_{\overline\square\square}\,
M_{\overline\square\square})$ of interest,
\begin{itemize}\item for the leading large-$N_{\rm c}$ behaviour of
the \emph{flavour-preserving\/} tetraquark-phile Green functions,
\begin{align*}\frac{\langle T(j_{\overline ab}\,j_{\overline cd}\,
j^\dag_{\overline ab}\,j^\dag_{\overline cd})\rangle}{f_M^4}&=
\frac{|A(T_A\longleftrightarrow M_{\overline ab}\,M_{\overline cd}
)|^2}{p^2-m^2_{T_A}}+\frac{|A(T_B\longleftrightarrow M_{\overline
ab}\,M_{\overline cd})|^2}{p^2-m^2_{T_B}}+\cdots=O(N_{\rm c}^{-2})
\ ,\\[1ex]\frac{\langle T(j_{\overline ad}\,j_{\overline cb}\,
j^\dag_{\overline ad}\,j^\dag_{\overline cb})\rangle}{f_M^4}&=
\frac{|A(T_A\longleftrightarrow M_{\overline ad}\,M_{\overline cb}
)|^2}{p^2-m^2_{T_A}}+\frac{|A(T_B\longleftrightarrow M_{\overline
ad}\,M_{\overline cb})|^2}{p^2-m^2_{T_B}}+\cdots=O(N_{\rm c}^{-2})
\ ,\end{align*}\item for the leading large-$N_{\rm c}$ behaviour
of the \emph{flavour-reordering\/} tetraquark-phile Green
functions,\begin{align*}\frac{\langle T(j_{\overline ad}\,
j_{\overline cb}\,j^\dag_{\overline ab}\,j^\dag_{\overline cd})
\rangle}{f_M^4}&=\frac{A(T_A\longleftrightarrow M_{\overline ab}\,
M_{\overline cd})\,A(T_A\longleftrightarrow M_{\overline ad}\,
M_{\overline cb})}{p^2-m^2_{T_A}}\\&+\frac{A(T_B
\longleftrightarrow M_{\overline ab}\,M_{\overline cd})\,A(T_B
\longleftrightarrow M_{\overline ad}\,M_{\overline cb})}
{p^2-m^2_{T_B}}+\cdots=O(N_{\rm c}^{-3})\ .\end{align*}
\end{itemize}For a single tetraquark, the conflict is evident. In
contrast, for two tetraquarks coupling unequally~to two ordinary
mesons, one solution for the large-$N_{\rm c}$ behaviour of the
four transition amplitudes reads
\setlength{\fboxsep}{2pt}\setlength{\fboxrule}{.5pt}\begin{align*}
\underbrace{A(T_A\longleftrightarrow M_{\overline ab}\,
M_{\overline cd})=O(N_{\rm c}^{-1})}_{\mbox{$\Longrightarrow\qquad
\Gamma(T_A)=O(N_{\rm c}^{-2})$}\ ,}\qquad\stackrel{\fbox{$N_{\rm
c}$ order}}{>}\qquad A(T_A\longleftrightarrow M_{\overline ad}\,
M_{\overline cb})=O(N_{\rm c}^{-2})&\\[1ex]A(T_B
\longleftrightarrow M_{\overline ab}\,M_{\overline cd})=O(N_{\rm
c}^{-2})\qquad\stackrel{\fbox{$N_{\rm c}$ order}}{<}\qquad
\underbrace{A(T_B\longleftrightarrow M_{\overline ad}\,
M_{\overline cb})=O(N_{\rm c}^{-1})}_{\mbox{$\Longrightarrow\qquad
\Gamma(T_B)=O(N_{\rm c}^{-2})$}\ ,}&\end{align*}if assuming both
tetraquark masses $m_{T_A}$ and $m_{T_B}$ to retain their finite
values in the limit \mbox{$N_{\rm c}\to\infty$. The} $N_{\rm c}$
dependence of the total decay widths $\Gamma(T_A)$ and
$\Gamma(T_B)$ of the two tetraquarks $T_A$ and $T_B$, governed by
their $N_{\rm c}$-leading two-meson channels, would be
parametrically equal: $\Gamma(T_A)=\Gamma(T_B)=O(N_{\rm c}^{-2})$.

\section{Bitter Pill to Swallow: Mismatch of Tetraquark Formation
and Large-$N_{\rm c}$ Insights}Finally, we turn to the formation
path of a \emph{compact\/} tetraquark meson. From the
group-theoretic point of view, a state combining two quarks and
two antiquarks transforming according to either~the 3-dimensional
fundamental representation $\mathbf{3}$ of the QCD gauge group
${\rm SU}(3)$ or that representation's complex conjugate
$\mathbf{\overline 3}$ forms a reducible 81-dimensional
representation of ${\rm SU}(3)$~whereas, according to confinement,
any hadron is an ${\rm SU}(3)$ singlet $\mathbf{1}$: we must
identify the singlets in the tensor~product$$q_b\,q_d\,\overline
q_a\,\overline q_c\sim
\mathbf{3\otimes3\otimes\overline{3}\otimes\overline{3}=81=1\oplus1
\oplus8\oplus8\oplus8\oplus8\oplus10\oplus \overline{10}\oplus27}\
.$$Tracking the tetraquark formation through the possible
intermediate stages, ${\rm SU}(3)$ singlets can~arise along only
two routes. Either the two quarks merge to a \emph{diquark\/} in
the antisymmetric 3-dimensional representation $\mathbf{\overline
3}$ and the two antiquarks to an \emph{antidiquark\/} in the
complex conjugate representation $\mathbf{3}$, which yields a
tightly bound, compact tetraquark state, or the supposed
tetraquark constituents~form two colour-singlet
\emph{quark--antiquark\/} states that may end up as a loosely
bound, molecular-type state. As a matter of fact, this reasoning
applies identically for all choices $N_{\rm c}\ge2$ of the number
of colours.

\section{Conclusion: Bad Hand for Tetraquark Mesons of Genuinely
Flavour-Exotic Type}So, large-$N_{\rm c}$ QCD causes, in $N_{\rm
c}$-leading evaluation, \emph{flavour-exotic\/} tetraquarks to
exist pairwise \cite{TQN1,TQP1,TQN2,TQN3} but provides just one
\emph{compact\/} tetraquark structure. The possible resolution of
that clash~might explain the lack of any reliable experimental
observation of flavour-exotic tetraquark~candidates~\cite{TQP2}:
\begin{quote}Large-$N_{\rm c}$ QCD does not support the existence
of any narrow \emph{flavour-exotic\/} tetraquarks.\end{quote}

Maybe, an evident \emph{side remark\/} is in order. For a pretty
simple reason, our above considerations do not apply to any case
of \emph{identical\/} flavours of either the two quarks or the two
antiquarks, and~thus also not to any case of identical flavours of
\emph{both\/} the two quarks and the two antiquarks, in
particular, not to any case of four identical flavours, that is,
to all tetraquarks exhibiting the flavour composition
\begin{equation}T=[\overline q_a\,q_b\,\overline q_c\,q_b]\quad
\mbox{($a,c$ arbitrary)}\qquad\mbox{or}\qquad T=[\overline
q_a\,q_b\,\overline q_a\,q_d]\quad\mbox{($b,d$ arbitrary)}\
:\label{fi}\end{equation}upon allowing for flavour identity $a=c$
or $b=d$ in Eq.~(\ref{fe}), flavour-retaining Green functions
(\ref{fpG}) and flavour-rearranging Green functions (\ref{frG})
become indistinguishable, whence a single tetraquark suffices to
satisfy the $N_{\rm c}$-leading constraints. Tetraquarks of the
kind (\ref{fi}) with diquark--antidiquark binding have been
explored since long, and may have been spotted by lattice QCD
recently \cite{FHLM,BSW,JMP}.

\vspace{3.39505ex}\noindent{\small{\bf Acknowledgements.\
}D.~M.~and H.~S.~express gratitude for support under joint
CNRS/RFBR Grant No.~PRC Russia/19-52-15022. D.~M.~is grateful for
support by the Austrian Science Fund (FWF), Project No.~P29028.}

\end{document}